\documentclass{article}
\usepackage[dvips]{color}
\usepackage{amsmath, amssymb, amsfonts}
\usepackage{epsfig}
\title{ Source with Nonzero Radial Pressure for the Vaidya Metric } 
\author{Hristu Culetu, \\Ovidius University, Department of Physics and Electronics, \\ Bld. Mamaia  124, 900527 Constanta, Romania \\e-mail: hculetu@yahoo.com}

\begin{document}
\numberwithin{equation}{section}
\pagenumbering{arabic}
\maketitle
\newcommand{\fv}{\boldsymbol{f}}
\newcommand{\tv}{\boldsymbol{t}}
\newcommand{\gv}{\boldsymbol{g}}
\newcommand{\OV}{\boldsymbol{O}}
\newcommand{\wv}{\boldsymbol{w}}
\newcommand{\WV}{\boldsymbol{W}}
\newcommand{\NV}{\boldsymbol{N}}
\newcommand{\hv}{\boldsymbol{h}}
\newcommand{\yv}{\boldsymbol{y}}
\newcommand{\RE}{\textrm{Re}}
\newcommand{\IM}{\textrm{Im}}
\newcommand{\rot}{\textrm{rot}}
\newcommand{\dv}{\boldsymbol{d}}
\newcommand{\grad}{\textrm{grad}}
\newcommand{\Tr}{\textrm{Tr}}
\newcommand{\ua}{\uparrow}
\newcommand{\da}{\downarrow}
\newcommand{\ct}{\textrm{const}}
\newcommand{\xv}{\boldsymbol{x}}
\newcommand{\mv}{\boldsymbol{m}}
\newcommand{\rv}{\boldsymbol{r}}
\newcommand{\kv}{\boldsymbol{k}}
\newcommand{\VE}{\boldsymbol{V}}
\newcommand{\sv}{\boldsymbol{s}}
\newcommand{\RV}{\boldsymbol{R}}
\newcommand{\pv}{\boldsymbol{p}}
\newcommand{\PV}{\boldsymbol{P}}
\newcommand{\EV}{\boldsymbol{E}}
\newcommand{\DV}{\boldsymbol{D}}
\newcommand{\BV}{\boldsymbol{B}}
\newcommand{\HV}{\boldsymbol{H}}
\newcommand{\MV}{\boldsymbol{M}}
\newcommand{\be}{\begin{equation}}
\newcommand{\ee}{\end{equation}}
\newcommand{\ba}{\begin{eqnarray}}
\newcommand{\ea}{\end{eqnarray}}
\newcommand{\bq}{\begin{eqnarray*}}
\newcommand{\eq}{\end{eqnarray*}}
\newcommand{\pa}{\partial}
\newcommand{\f}{\frac}
\newcommand{\FV}{\boldsymbol{F}}
\newcommand{\ve}{\boldsymbol{v}}
\newcommand{\AV}{\boldsymbol{A}}
\newcommand{\jv}{\boldsymbol{j}}
\newcommand{\LV}{\boldsymbol{L}}
\newcommand{\SV}{\boldsymbol{S}}
\newcommand{\av}{\boldsymbol{a}}
\newcommand{\qv}{\boldsymbol{q}}
\newcommand{\QV}{\boldsymbol{Q}}
\newcommand{\ev}{\boldsymbol{e}}
\newcommand{\uv}{\boldsymbol{u}}
\newcommand{\KV}{\boldsymbol{K}}
\newcommand{\ro}{\boldsymbol{\rho}}
\newcommand{\si}{\boldsymbol{\sigma}}
\newcommand{\thv}{\boldsymbol{\theta}}
\newcommand{\bv}{\boldsymbol{b}}
\newcommand{\JV}{\boldsymbol{J}}
\newcommand{\nv}{\boldsymbol{n}}
\newcommand{\lv}{\boldsymbol{l}}
\newcommand{\om}{\boldsymbol{\omega}}
\newcommand{\Om}{\boldsymbol{\Omega}}
\newcommand{\Piv}{\boldsymbol{\Pi}}
\newcommand{\UV}{\boldsymbol{U}}
\newcommand{\iv}{\boldsymbol{i}}
\newcommand{\nuv}{\boldsymbol{\nu}}
\newcommand{\muv}{\boldsymbol{\mu}}
\newcommand{\lm}{\boldsymbol{\lambda}}
\newcommand{\Lm}{\boldsymbol{\Lambda}}
\newcommand{\opsi}{\overline{\psi}}
\renewcommand{\tan}{\textrm{tg}}
\renewcommand{\cot}{\textrm{ctg}}
\renewcommand{\sinh}{\textrm{sh}}
\renewcommand{\cosh}{\textrm{ch}}
\renewcommand{\tanh}{\textrm{th}}
\renewcommand{\coth}{\textrm{cth}}

\begin{abstract}
A null fluid with radial pressure is proposed as a source generating Vaidya spacetime. The fluid is anisotropic with no transversal pressures and $p = \rho/3$ as its equation of state, where $p$ is the isotropic pressure and $\rho$ is the energy density. The radial energy flux is directed inward when $\dot{m} > 0$ and the total energy flow crossing a surface of constant $r$ resembles the Brown--York quasilocal energy (QLE). A modified Vaidya metric removes the singularity at the origin, and the corresponding QLE is investigated in terms of both $r$ and the mass $m(v)$ of the object.
 \end{abstract}
 
 \section{.Introduction}
  In general relativity, the Vaidya solution$^{1)}$ is a nonstatic generalization of the Schwarzschild geometry and has some unique features. It describes the geometry of unpolarized radiation, represented by a null fluid emerging from a spherically symmetric source. Misner $^{2)}$ considered a sphere of fluid subject to gravitational and pressure forces, where each element of the fluid is cooled by the radially outward emission of neutrinos. He matched the fluid sphere with an exterior region given by the ,,radiating Schwarzschild'' metric found by Vaidya$^{1)}$. In a subsequent paper, Lindquist et al.$^{3)}$ showed that in Vaidya's spacetime for a radiating sphere,$-dm/du$ (where $m$ is the sphere mass and $u$ is the retarded time coordinate) is the total power output as given by the Landau-Lifshitz energy-momentum pseudotensor. They also studied the geodesics in the Vaidya geometry and found that the acceleration $a_{L} = -L/r$ is a non-Newtonian gravitational field (induction field) associated with the luminosity $L$ of the central source.
	
	Fayos et al.$^{4)}$ found the necessary conditions to perform the matching (across a three-dimensional timelike surface with spherical symmetry) of the Robertson--Walker metric and radiating Vaidya metric (see also Ref.5), where the authors stated that every spherically symmetric metric may be locally matched to an exterior Vaidya solution provided that there is a timelike surface such that the total radial pressure vanishes on it).

	The end state of the collapse of null radiation with a string fluid was considered by Govinder and Govender$^{6)}$. They also investigated the collapse of Ricci-flat metrics and showed that the presence of a string fluid could lead to the appearance of a locally naked singularity for null fluids. McClure et al.$^{7)}$ analyzed cosmological versions of Vaidya's radiating stellar exterior. They found a two-fluid solution that consists of a null fluid and an imperfect fluid that contains an inhomogeneous dark energy (DE) component with negative energy density for the spacetimes of an accelerating cosmological reference frame.
  
  The black hole (BH) evaporation in conformal gravity was studied by Bambi et al.$^{8)}$. Since massless particles are naturally conformally invariant, the simplest model of gravitational collapse in conformal gravity is given by the collapse of a thin shell of radiation with Vaidya's space outside the shell that is conformally flat inside. They chose the radiation stress tensor to be that of a null dust. 
  
  Our motivation in this paper is to look for a null fluid with nonzero radial pressure as the source of the Vaidya geometry. Section 2 introduces the nonperfect fluid stress tensor and its structure, and the kinematical properties of a timelike congruence of observers are studied. In Section 3  the total energy flow through an $r = const.$ surface and its relation with the Brown--York energy $W$ are calculated, where $r$ is the radial coordinate. A modified regular form of the Vaidya metric is introduced in Section 4 and its properties are investigated. The modified quasilocal energy $W$ is analyzed in Section 5, both as a function of $r$ and as a function of mass $m$ of the object, a celestial body, or a BH. We conclude with some general remarks in Section 6. Throughout the paper we use geometrical units $G = c = 1$, unless otherwise specified.
  
  \section{.Null Fluid with Radial Pressure}
  Vaidya$^{9)}$ solved Einstein's equations for a spherically symmetric radiative body with the stress tensor of radiation $T_{ab} = \rho l_{a} l_{b}$, where $l_{a}$ is a null vector directed radially and $\rho$ is defined as the energy density of the incoming radiation measured locally by an observer with 4-velocity field $u^{a}$, so that$^{10)}$ $ \rho = T_{ab} u^{a} u^{b}$.
   The line element can be written as
     \begin{equation}
  ds^{2} = -(1 - \frac{2m(v)}{r})dv^{2} + 2dvdr + r^{2} d \Omega^{2}, 
 \label{2.1}
 \end{equation}
where $v$ is the advanced ingoing time coordinate and $r$ is the radial coordinate. The parameter $m(v)$ is the increasing mass of the radiating spherically symmetric object and $d \Omega^{2}$ stands for the metric of the unit two-sphere. We consider only the region $r > 2m(v)$ for Eq. (2.1) to avoid the signature flip at the horizon (see also Ref. 2). It is valid for both ordinary celestial objects and for BHs. The geometry given by Eq. (2.1) is generated by the energy-momentum tensor given above, which has only one nonzero component $T^{r}_{~v} = \dot{m}/4\pi r^{2}$ and zero trace, where $\dot{m} = dm/dv > 0$. In other words, it corresponds to a null fluid with no pressures, i.e., a null dust. This interpretation originates from the fact that the tensor 
     \begin{equation}
	T^{a}_{~b} = \rho l^{a} l_{b}	
 \label{2.2}
 \end{equation}
resembles the stress tensor for a perfect fluid 
     \begin{equation}
		  T_{ab} = (p + \rho) u_{a} u_{b} + p g_{ab} 
 \label{2.3}
 \end{equation}
when $p = 0$. 

We consider now a different interpretation of the source of the Vaidya metric and take the energy-momentum tensor corresponding to a nonperfect fluid with energy flux$^{10)}$
     \begin{equation}
		  T_{ab} = (p_{t} + \rho) u_{a} u_{b} + p_{t} g_{ab} + (p_{r} - p_{t}) n_{a}n_{b} +  u_{a} q_{b} + u_{b} q_{a},
 \label{2.4}
 \end{equation}
where $p_{r}$ is the radial pressure, $p_{t}$ are the transversal (tangential) pressures, $q^{a}$ is the energy flux 4-vector, and $n^{a}$ is a unit spacelike vector orthogonal to $u^{a}$. We have $u_{a} n^{a} = 0,~u_{a} u^{a} = -1,~n_{a} n^{a} = 1$, and $u_{a} q^{a} = 0$. We look for expressions of the quantities $\rho,~p_{r},~ p_{t},~q^{a}$ in Eqs. (2.2) and (2.4) to represent the same stress tensor. 

Let us take a velocity vector field of the form [for simplicity we replace $m(v)$ with $m$]
     \begin{equation}
	 u^{a} = \left(\frac{1}{\sqrt{1 - \frac{2m}{r}}}, 0, 0, 0 \right),	
 \label{2.5}
 \end{equation}
i.e., an observer sitting at $r = const.$ From Eq. (2.5) and the properties of $n^{a}$, we find that 
     \begin{equation}
	 n^{a} = \left(\frac{1}{\sqrt{1 - \frac{2m}{r}}},\sqrt{1 - \frac{2m}{r}} , 0, 0 \right),~~~	n_{a} = \left(0, \frac{1}{\sqrt{1 - \frac{2m}{r}}}, 0, 0 \right).	
 \label{2.6}
 \end{equation}
As we previously specified, the mixed stress tensor that generates Eq. (2.1) has only one nonzero component: $T^{r}_{~v} = \dot{m}/4\pi r^{2}$. Defining, as is obvious, the energy density of the null fluid as $ \rho = T^{a}_{~b} u^{b} u_{a}$, it is an easy task to obtain
     \begin{equation}
		\rho = \frac{\dot{m}}{4\pi r^{2} (1 - \frac{2m}{r})},
 \label{2.7}
 \end{equation}
where $u_{a} = (- \sqrt{1 - \frac{2m}{r}}, 1/\sqrt{1 - \frac{2m}{r}}, 0, 0)$ has been used. Keeping in mind that $T^{\theta}_{~\theta} = T^{\phi}_{~\phi} = 0$, the expression for $p_{t}$ results immediately from (2.4): $p_{t} = 0$, so that the fluid has no tangential pressures. The trace of (2.4) yields
     \begin{equation}
	 T^{a}_{~a} = -\rho + p_{r} + 2p_{t} = 0,	
 \label{2.8}
 \end{equation}
whence $ p_{r} = \rho$. Hence, the null fluid has a nonzero radial pressure (the fact that $ p_{r} = \rho$ and $ p_{t} = 0$ resembles the stress tensor for a directed flow of radiation$^{11)}$). From here we find that the isotropic pressure is $p = (p_{r} + 2p_{t})/3 = p_{r}/3 = \rho/3$, as for incoherent radiation$^{6)}$ (see also Ref. 1). In addition, it reduces to a dissipative fluid with a stiff equation of state (the radial speed of sound equals the speed of light). This is not surprising because the fluid behaves as the directed flow of radiation described by Tolman$^{11)}$ (in our case the direction is the radial one). From the energy flux 4-vector (see below), one finds that the invariant radial flux is $q^{a}n_{a} = -\rho$, as expected for ingoing radiation.

  The current is obtained from (2.4) as
     \begin{equation}
	q^{a} = - T^{a}_{~b}u^{b} - \rho u^{a},	
 \label{2.9}
 \end{equation}
which gives us
     \begin{equation}
	q^{a} = \left(-\frac{\dot{m}}{4\pi r^{2}(1 - \frac{2m}{r})^{3/2}}, -\frac{\dot{m}}{4\pi r^{2}(1 - \frac{2m}{r})^{1/2}}, 0 ,0\right)
 \label{2.10}
 \end{equation}
and $q = \sqrt{q^{a} q_{a}} = |\rho|$, as expected for a null fluid. 

Naw having the physical parameters in Eq. (2.4), we may compare  $T^{a}_{~b}$ from Eq. (2.4) with that from Eq. (2.2) to obtain 
     \begin{equation}
	l^{a} = u^{a} - n^{a},
 \label{2.11}
 \end{equation}
whence
     \begin{equation}
	l^{a} = (0, -\sqrt{1 - \frac{2m}{r}}, 0, 0),~~~ l_{a} = ( -\sqrt{1 - \frac{2m}{r}}, 0, 0, 0).
 \label{2.12}
 \end{equation}
Using $u^{a}$ from Eq. (2.5), we obtain the components of the acceleration 4-vector of the congruence as
     \begin{equation}
	a^{v} = \frac{m}{r^{2} (1 - \frac{2m}{r})} + \frac{\dot{m}}{r (1 - \frac{2m}{r})^{2}},~~~a^{r} =  (1 - \frac{2m}{r}) a^{v},~~~a^{\theta} = a^{\phi} = 0. 
 \label{2.13}
 \end{equation}
The proper acceleration reads
     \begin{equation}
\sqrt{a^{b}a_{b}}	= \frac{\dot{m}}{ r(1 - \frac{2m}{r})^{3/2}} + \frac{m}{ r^{2}(1 - \frac{2m}{r})^{1/2}}	
 \label{2.14}
 \end{equation}
and the scalar expansion is given by
     \begin{equation}
		\Theta \equiv \nabla_{a}u^{a} = \frac{\dot{m}}{r (1 - \frac{2m}{r})^{3/2}}.
 \label{2.15}
 \end{equation}
 In the standard Vaidya solution the radial pressure does not appear. In our interpretation, it is hidden by the null vector $l^{a}$ which is a difference between a timelike vector $u^{a}$ and a spacelike vector $n^{a}$ that is orthogonal to $u^{a}$. The standard interpretation of a null dust is introduced by a formal analogy with a perfect fluid with no pressures. We conjectured a general expression [Eq. (2.4)] for the source and forced it to be traceless and to have only one nonzero component, i.e., $T^{r}_{~v} $, in accordance with the standard solution. 

We set $r > 2m(v)$, otherwise $v$ becomes a spacelike coordinate (see also Ref. 3). With $\dot{m} > 0$ (we are considering ingoing radiation), $\rho = p_{r} \geq 0$ ($\rho = 0$ is obtained when $\dot{m} = 0$, namely in the static situation). In other words, the weak energy condition (WEC) ($\rho \geq 0,~\rho + p_{r} \geq 0,~\rho + p_{t} \geq 0$) is obeyed. This is valid for other energy conditions: the null energy condition (NEC) ($\rho + p_{r} \geq 0,~\rho + p_{t} \geq 0)$, strong energy condition (SEC) ($\rho + p_{r} \geq 0,~\rho + p_{t} \geq 0,~\rho + p_{r} + 2p_{t} \geq 0$) and dominant energy condition (DEC) ($\rho > |p_{r}|,~ \rho > |p_{t}|$). Being null, the fluid satisfies the condition $\rho = |q|$. Note that $\rho$ is divergent at the origin and at the evolutionary hypersurface $r = 2m(v)$ (the apparent horizon) because of the nonstatic character of the spacetime. The location of the apparent horizon may be obtained from $g^{ab}R_{,a}R_{,b} = 0$, where $R = r$ is the areal radius. One obtains $g^{rr} = 1 - 2m(v)/r = 0$ or $r = 2m(v)$. We could have also used the method from Ref. 12 ( the vanishing of the scalar expansion of the null vector field along null geodesics). 

We shall see later how the divergence at $r = 0$ can be removed. The currents $q^{r}$ and $q^{v}$ from Eq. (2.10) are directed inward because $\dot{m} > 0$ and are also divergent on the horizon. As far as the acceleration of the congruence is concerned, we notice that the invariant acceleration from (2.14) becomes the standard Schwarzschild counterpart when $\dot{m} = 0$. 

\section{.Brown--York Quasilocal Energy}
Now having the components of the stress tensor and the basic physical quantities associated with it, our next task is to compute the total energy flow measured by an observer located at $r = const.$$^{13)}$. It is given by
 \begin{equation}
 W = \int{T^{a}_{~b}u^{b}n_{a}\sqrt{-\gamma}}dv~d\theta~d\phi,
 \label{3.1}
 \end{equation}
where $u^{b}$ and $n^{b}$ are given by Eq. (2.5) and Eq. (2.6), respectively. To find $\gamma$ (i.e., the determinant of the induced metric on the hypersurface of $r = const.$) we take $dr = 0$ in (2.1) and obtain $\gamma = -(1 - 2m/r)r^{4} sin^{2}\theta$ (see also Ref. 14). With $T^{r}_{~v} = \dot{m}/4\pi r^{2}$, Eq. (3.1) yields
 \begin{equation}
 W = \int{T^{r}_{~v}u^{v}n_{r}\sqrt{-\gamma}}dv~d\theta~d\phi = \int{\frac{dm}{\sqrt{1 - \frac{2m}{r}}}},
 \label{3.2}
 \end{equation}
where $r$ is fixed and $m$ is the variable of integration. One obtains
 \begin{equation}
 W(m) = -r \sqrt{1 - \frac{2m}{r}} +g(r),
 \label{3.3}
 \end{equation}
with $g(r)$ a free function of $r$. $g(r)$ may be determined by imposing $W = 0$ when $m$ vanishes, so that $g(r) = r$. Hence,
 \begin{equation}
 W(m) = r \left(1 -  \sqrt{1 - \frac{2m}{r}}\right).
 \label{3.4}
 \end{equation}
 A plot of $W(m)$ versus $m$ at constant $r$ is depicted in Fig. 1. Note that the derivative $\partial W/\partial m$ diverges at $m = r/2$. Equation (3.4), taken as a function of $r$, resembles the quasilocal energy (QLE) for a Schwarzschild BH$^{15)}$. Note that Lundgren et al. plotted $W(r)$ even for $r < 2m$. However, we have to keep in mind that $r$ becomes timelike in this region, and covering both regions ($r < 2m$ and $r > 2m$) in the same plot seems to be inappropriate. 

Because of the square root in Eq. (3.4), the derivative $\partial W/\partial m$ is infinite at the horizon, which is similar to the infinite value of $\partial E(r)/\partial r$ from the approach of Lundgren et al. We also observe that the derivation of the QLE $E(r)$ by Lundgren et al.$^{15)}$ remains valid even when the mass of the object is time-dependent, as in our situation, because the connection coefficients that we need have the same expressions as in the static case. Therefore, the extrinsic curvature of the two-boundary preserves its value.

\section{.Regularized Vaidya Spacetime}

 The Vaidya metric has the same singularity at the origin $r = 0$ as the Schwarzschild metric. Bardeen$^{16)}$ first presented a regular BH model. However, the physical source associated with his solution was clarified much later when Ayon-Beato and Garcia$^{17)}$ interpreted it as the gravitational field of a nonlinear magnetic monopole of a self-gravitating magnetic field. In the framework of loop quantum gravity, Gambini and Pullin$^{18)}$ eliminated the singularity, replacing it by a region of high curvature, yielding a global structure similar to that of the Reissner-Nordstrom spacetime but without singularities.

It is clear from Eq. (2.1) that we must impose $r \geq 2m(v)$ since at $r = 2m(v)$ a signature flip takes place and $v$ becomes a spacelike coordinate. In addition, being nonstatic, the metric in Eq. (2.1) has no timelike Killing vector, and so an event horizon cannot be obtained. In this case, an apparent horizon is more appropriate. To extend the metric in Eq. (2.1) beyond the evolutionary surface $r = 2m(v)$, we introduce an exponential factor in the metric, rendering it and the stress tensor regular at the origin. Therefore, we propose the geometry 
 \begin{equation}
   ds^{2} = -\left(1 - \frac{2m(v)}{r} e^{-\frac{2m(v)}{er}}\right) dv^{2} + 2dvdr + r^{2} d \Omega^{2}     
 \label{4.1}
 \end{equation}
with $lne =1$. The above line element may be obtained in a similar manner to that for Eq. (2.1). We consider the regularized Schwarzschild spacetime$^{19)}$
 \begin{equation}
      ds^{2} = -\left(1 - \frac{2m}{r} e^{-\frac{2m}{er}}\right) dt^{2} + \frac{1}{1 - \frac{2m}{r} e^{-\frac{2m}{er}}} dr^{2} + r^{2} d \Omega^{2}         
 \label{4.2}
 \end{equation}
with $k$ (from Ref. 19) = $2m/e$. It is clear that (4.2) is regular at the origin (more precisely, it is Minkowskian there). In addition, the stress tensor is regular everywhere. When a BH has charge $q$ and $k = q^{2}/2m$, the energy-momentum tensor acquires exactly the Maxwell stress tensor form in the region $r >> k$. This justifies that the stress tensor generating the geometry (4.2) represents a physical matter source.  

From Eq. (4.2) we move to Eddington--Finkelstein-type coordinates with 
 \begin{equation}
dr^{*} = \frac{dr}{1 - \frac{2m}{r} e^{-\frac{2m}{er}}},~~~v = t + r^{*} 
 \label{4.3}
 \end{equation}
and obtain the metric (4.1) but with $m = const.$ We then simply take $m = m(v)$ and arrive at the line-element (4.1). 

It was shown in Ref. 19 that, for $m = const., -g_{vv} \equiv f(r,m)$ from (4.1) is nonnegative for any $r > 0$ and vanishes at $r = 2m/e \equiv r_{H}$. We now check the positivity of $-g_{vv}$ for any $m(v) > 0$ with $r = const.$ We have
 \begin{equation}
\frac{\partial f}{\partial m} = -\frac{2}{r} \left(1 - \frac{r_{H}}{r}\right)e^{-\frac{r_{H}}{r}},
 \label{4.4}
 \end{equation}
which vanishes at $m = er/2$ and $f(er/2) = 0 = f_{min}$. Therefore, $f(m)$ is nonnegative for any positive $m$. A plot of $f(m)$ against $m$ for constant $r$ is given in Fig. 2. In addition, $f(m)$ tends to unity at $m = 0$ and when $m \rightarrow \infty$. The dependence of $f(m)$ on $m$ is similar to that of $f(r)$ at constant $m$$^{19)}$. Hence, $0 \leq f(r,m) \leq 1$ for any $r$ and $m$. 

As far as the components of $u^{a}$ and $n^{a}$ in Eq. (4.1) are concerned, they have the same structure as those from Eq. (2.5) and Eq. (2.6) with the exception of the square root, which is $(1 - \frac{2m(v)}{r} e^{-\frac{2m(v)}{er}})^{1/2}$. From the expression for $u^{a}$, one obtains the radial acceleration of the congruence as
 \begin{equation}
a^{r} =   \frac{m(1 - \frac{r_{H}}{r})e^{-\frac{r_{H}}{r}}}{r^{2}} + \frac{\dot{m}(1 - \frac{r_{H}}{r})e^{-\frac{r_{H}}{r}}}{r \left(1 - \frac{2m}{r} e^{-\frac{r_{H}}{r}}\right)}. 
 \label{4.5}
 \end{equation}
One notices that $a^{r}$ becomes negative when $r < r_{H}$ and vanishes on the horizon. Hence, the system has repulsive properties inside of the BH. We also observe that the second term from the r.h.s. of (4.5) has the same sign as the first term because $\dot{m}$ is positive.

 We now obtain the following expressions for $\rho$ and $p_{r}$ using the same arguments as those from Sec. 2:
 \begin{equation}
8\pi \rho = \frac{4m^{2}}{ er^{4}} e^{-\frac{2m}{er}} +  \frac{2\dot{m}(1 - \frac{r_{H}}{r})e^{-\frac{r_{H}}{r}}}{r^{2} \left(1 - \frac{2m}{r} e^{-\frac{r_{H}}{r}}\right)} 
 \label{4.6}
 \end{equation}
and
 \begin{equation}
8\pi p_{r} = -\frac{4m^{2}}{ er^{4}} e^{-\frac{2m}{er}} +  \frac{2\dot{m}(1 - \frac{r_{H}}{r})e^{-\frac{r_{H}}{r}}}{r^{2} \left(1 - \frac{2m}{r} e^{-\frac{r_{H}}{r}}\right)}.   
 \label{4.7}
 \end{equation}
The transversal pressures are
 \begin{equation}
8\pi p_{t} = \frac{4m^{2}}{ er^{4}} \left(1 - \frac{m}{er}\right) e^{-\frac{2m}{er}},   
 \label{4.8}
 \end{equation}
which do not depend on $\dot{m}$. Note that the above quantities may be neglected at large radii ($r >> r_{H}$), where all the energy conditions are satisfied. They also vanish at $r = 0$ due to the exponential factor. In contrast, $ p_{r}$ is always negative for $r < r_{H}$ but the sign of $\rho $ in the same region depends on the value of $\dot{m}$. In addition, the tangential pressures $ p_{t}$ become negative when $r < r_{H}/2$. In conclusion, the energy conditions are not satisfied in the region $r < r_{H}$.

The energy flux 4-vector has the nonzero components
 \begin{equation}
q^{v} = -  \frac{\dot{m}\left(1 - \frac{r_{H}}{r}\right)e^{-\frac{r_{H}}{r}}}{4\pi r^{2} \left(1 - \frac{2m}{r} e^{-\frac{r_{H}}{r}}\right)\sqrt{1 - \frac{2m}{r} e^{-\frac{r_{H}}{r}}}},~~~ q^{r} = \left(1 - \frac{2m}{r} e^{-\frac{r_{H}}{r}}\right) q^{v},
 \label{4.9}
 \end{equation}
with
 \begin{equation}
\sqrt{q^{a}q_{a}} =  \frac{\dot{m}|1 - \frac{r_{H}}{r}|e^{-\frac{r_{H}}{r}}}{4\pi r^{2} \left(1 - \frac{2m}{r} e^{-\frac{r_{H}}{r}}\right)} .  
 \label{4.10}
 \end{equation}
One observes that the trace
 \begin{equation}
T^{a}_{~a} = -\frac{8m^{3}}{ e^{2}r^{5}} e^{-\frac{2m}{er}}  
 \label{4.11}
 \end{equation}
is always negative. Therefore, we no longer have a null fluid but an anisotropic fluid with energy flux. In the static case ($\dot{m} = 0$), $\rho$ and $p_{r}$ acquire the corresponding values from Ref. 19, which are regular in the whole spacetime. By comparing with $\rho$ (and $p_{r}$) in Eq. (2.7), it can be seen that they vanish at the origin. The conclusion is that the exponential factor from the line element (4.1) renders $\rho$ and $p_{r}$ finite only at the origin of the coordinates and not at the apparent horizon $r = r_{H}$owing to the variable mass $m(v)$. In addition, the nonstatic character of the metric in (4.1) leads to a nonvanishing scalar expansion for the congruence,
 \begin{equation}
		\Theta \equiv \nabla_{a}u^{a} =  \frac{\dot{m}(1 - \frac{r_{H}}{r})e^{-\frac{r_{H}}{r}}}{r \left(1 - \frac{2m}{r} e^{-\frac{r_{H}}{r}}\right)^{3/2}}, 
 \label{4.12}
 \end{equation}
which diverges at $r = r_{H}$, is negative for $r < r_{H}$, positive for $r > r_{H}$, and vanishes when $r \rightarrow 0$. 

\section{.Brown--York Energy for the Modified Vaidya Metric}
To find an expression for the total energy $W$ crossing a surface of constant $r$, we follow the same prescription as that from the beginning of Sec. 3. We now obtain 
 \begin{equation}
T^{r}_{~v}  =   \frac{\dot{m}(1 - \frac{r_{H}}{r})e^{-\frac{r_{H}}{r}}}{4\pi r^{2}} ,
 \label{5.1}
 \end{equation}
and
\begin{equation}
u^{v} = n_{r} = 1/\sqrt{1 - \frac{r_{H}}{r}e^{-\frac{r_{H}}{r}}},~~~ \sqrt{-\gamma} = \sqrt{1 - \frac{r_{H}}{r}e^{-\frac{r_{H}}{r}}} r^{2}sin\theta.
\label{5.2}
\end{equation}
 Therefore,
 \begin{equation}
W = \int{ \frac{\dot{m}(1 - \frac{r_{H}}{r})e^{-\frac{r_{H}}{r}}}{ r^{2}\left(1 - \frac{2m}{r} e^{-\frac{r_{H}}{r}}\right)} \sqrt{1 - \frac{2m}{r} e^{-\frac{r_{H}}{r}}}r^{2}}dv,
 \label{5.3}
 \end{equation}
which gives us
 \begin{equation}
W = \int{ \frac{(1 - \frac{r_{H}}{r})e^{-\frac{r_{H}}{r}}}{ \sqrt{1 - \frac{2m}{r} e^{-\frac{r_{H}}{r}}}}}~dm .
 \label{5.4}
 \end{equation}
We finally obtain
 \begin{equation}
 W(m) = r \left(1 -  \sqrt{1 - \frac{2m}{r} e^{-\frac{r_{H}}{r}}}\right),
  \label{5.5}
	\end{equation}
 where the same boundary conditions have been applied, as in Eq. (3.4). Note that $W(m)$ seems to have the same mathematical structure as Eq. (3.4), although the integrands were very different. This gives us the opportunity to consider the same physical meaning as in Sec. 3: $W(m)$ represents the Brown--York energy enclosed by an $r = const.$ hypersurface.

Let us first study the behavior of $W$ from (5.5) w.r.t. $r$, taking $m = const.$ We observe that $W(r) \rightarrow m$ when $r \rightarrow \infty$ , as expected, and vanishes at $r = 0$. We might, of course, calculate $W(r)$ directly, following the recipe from Ref. 15,
 \begin{equation}
 W(r) = \frac{1}{8\pi} \int_{B}{(K - K_{0})\sqrt{\sigma}d^{2}x},
  \label{5.6}
	\end{equation}
where $\sigma = det(\sigma_{ab})$ and $\sigma_{ab} = g_{ab} + u_{a}u_{b} - n_{a}n_{b}$ is the induced metric on the two-boundary $B$. $K$ in Eq. (5.6) is the trace of the extrinsic curvature of $B$ and $K_{0}$ corresponds to the flat Minkowski spacetime (i.e., when $m = 0$). However, the connection coefficients in Eq. (3.5) are also valid for the spacetime in Eq. (4.1) so that Eq. (5.5) emerges. 

A more detailed investigation of the function $W(r)$ requires the derivative $\partial W/\partial r$. We now look for the extremal of $W(r)$, if any. From
 \begin{equation} 
\frac{\partial W(r)}{\partial r} = 1 -  \sqrt{1 - \frac{2m}{r} e^{-\frac{r_{H}}{r}}} - \frac{m(1 - \frac{r_{H}}{r})e^{-\frac{r_{H}}{r}}}{r \sqrt{1 - \frac{2m}{r} e^{-\frac{r_{H}}{r}}}} = 0, 
  \label{5.7}
	\end{equation}
 after squaring, we arrive at
 \begin{equation}
1 + \frac{r_{H}}{r} = \frac{2}{\sqrt{e}} e^{\frac{r_{H}}{2r}},
  \label{5.8}
	\end{equation}
which is a transcendental equation. We show that (5.8) has a unique solution. A new variable $ x = r_{H}/2r$ yields
 \begin{equation}
e^{x} = \frac{\sqrt{e}}{2} (1 + 2x).
  \label{5.9}
	\end{equation}
One observes that the two curves (the straight line from the r.h.s. and the exponential) have the same tangent at $x = 1/2$,  where the two functions are equal. In conclusion, $r = r_{H}$ is the single solution of (5.8). Nevertheless, at $r = r_{H},~W(r)$ is not extremal because the denominator vanishes there. Taking the limit of the derivative when $r \rightarrow r_{H}$, one obtains the value $1 + \sqrt{2}/2$ when $r < r_{H}$ and $1 - \sqrt{2}/2$ for $r > r_{H}$. However, we see that at the apparent horizon $r = r_{H},~\partial W/\partial r$ is  finite, contrary to its infinite value from Ref. 15 [see the authors' comments above their Eq. (18)]. We also have $W(r_{H}) = r_{H} = 2m/e$. Fig. 3 depicts the QLE in terms of $r$ both inside and outside the horizon. 

Let us now look at the function $W(m)$ given by Eq. (5.5) at constant $r$. Note that $W(m)$ is undetermined as we do not have an expression for $m(v)$. Our procedure gives no indication of how to determine this expression. As in the previous section, one must solve the equation $\partial W/\partial m = 0$ to find the extremal values of $W(m)$, if any. 
From (5.5) we have
 \begin{equation}
\frac{\partial W(m)}{\partial m} =  \frac{(1 - \frac{r_{H}}{r})e^{-\frac{r_{H}}{r}}}{ \sqrt{1 - \frac{2m}{r} e^{-\frac{r_{H}}{r}}}}.  
  \label{5.10}
	\end{equation}
Therefore, $W(m)$ increases from zero at $m = 0$ to $r$ at $m = er/2$ in the region $m < er/2$ and decreases from $r$ to zero at infinity in the region $m > er/2$ for a given $r$. The derivative in Eq. (5.10) is not defined at $m = er/2$, where both the numerator and denominator vanish. Therefore, we have to compute the side limits at this point, which give the values $\sqrt{2}/e$ for $m < er/2$ and $-\sqrt{2}/e$ for $m > er/2$. In other words, $W(m)$ is not differentiable at $m = er/2$ but both side limits are finite. Figure 4 shows the dependence of $W(m)$ on $m$. There is an inflexion point at $m = er$. Consequently, $W(m)$ is always positive, vanishes at $m = 0$, and acquires its maximum value when the observer is located at the horizon.

\section{.Concluding Remarks}
The properties of the Vaidya solution of Einstein's equations with a null fluid as the source were investigated in this paper. In contrast with the standard Vaidya solution, our fluid has a nonzero radial pressure equal to the energy density. In our region of interest $r > 2m(v)$, all energy conditions for the stress tensor are satisfied. Owing to the vanishing tangential pressures, the equation of state of the fluid is $p = \rho/3$, in agreement with that of null radiation. The total energy flow crossing an $r =const.$ hypersurface turns out to represent the Brown--York quasilocal energy.

\begin{figure}
\includegraphics[width=4.0cm]{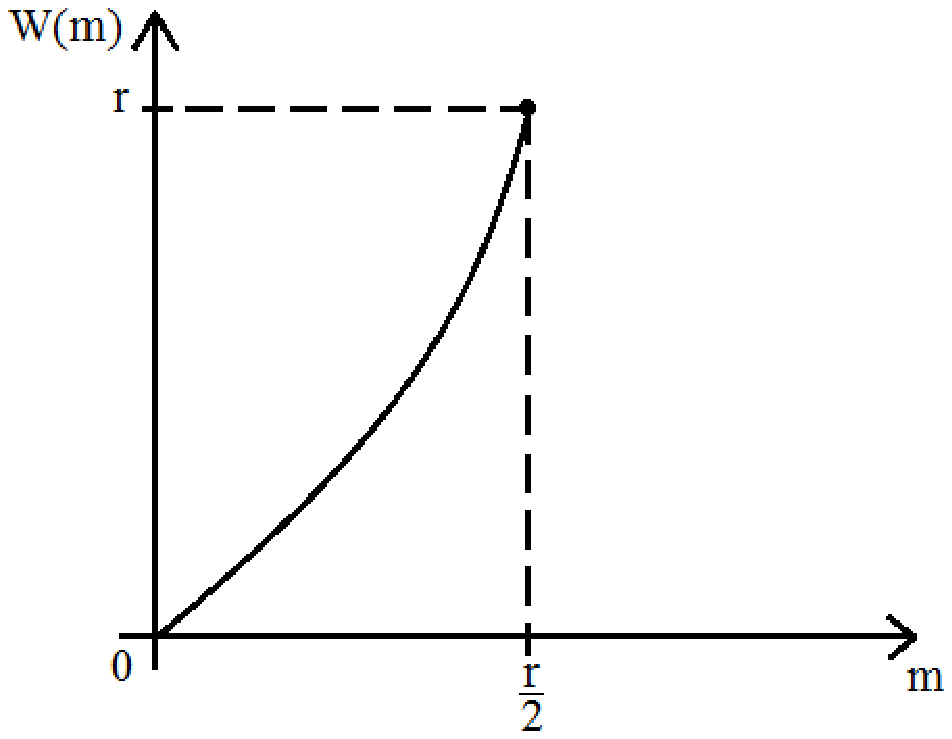}
\caption{Total energy flow measured by an observer located at $r = const.$. $W(m)$ is not differentiable at $m = r/2$.}
\label{68389Fig1}
 \includegraphics[width=4.0cm]{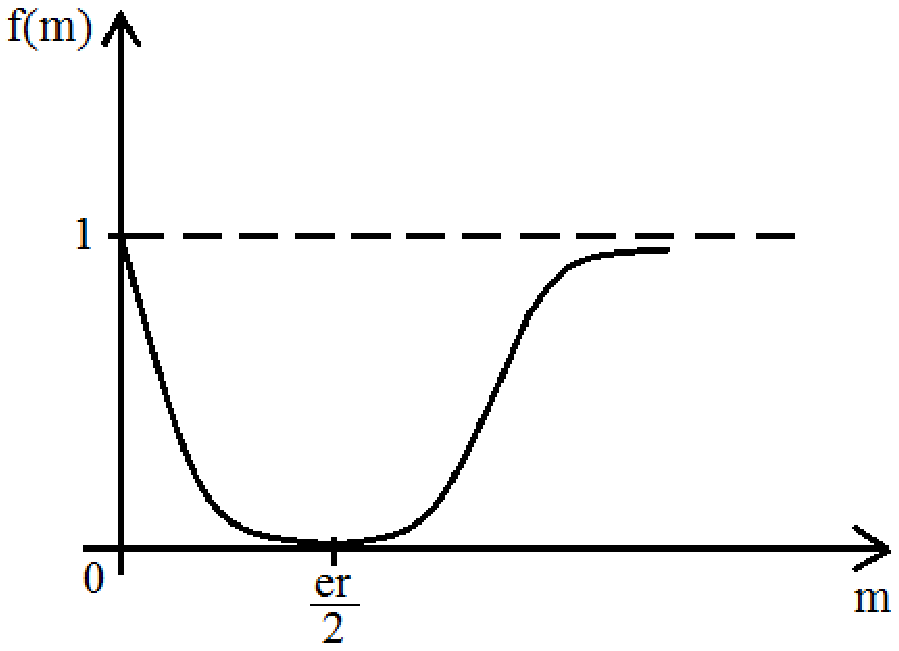}
\caption{Metric coefficient $-g_{vv} = f(m)$ from Eq. (4.4) for constant $r$.}
\label{68389Fig2}
\end{figure}
\begin{figure}
\includegraphics[width=4.0cm]{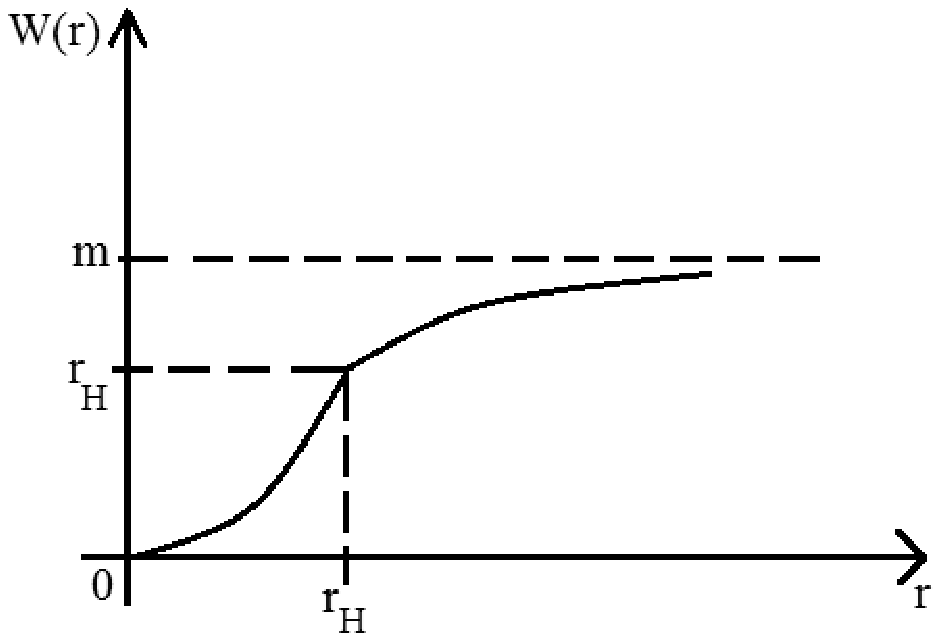}
\caption{Quasilocal energy $W(r)$ from Eq. (5.5) versus $r$ at constant $v$. $W(r)$ is not differentiable at the horizon $r = r_{H}$ and becomes the ADM mass at infinity.}
\label{68389Fig3}
\includegraphics[width=4.0cm]{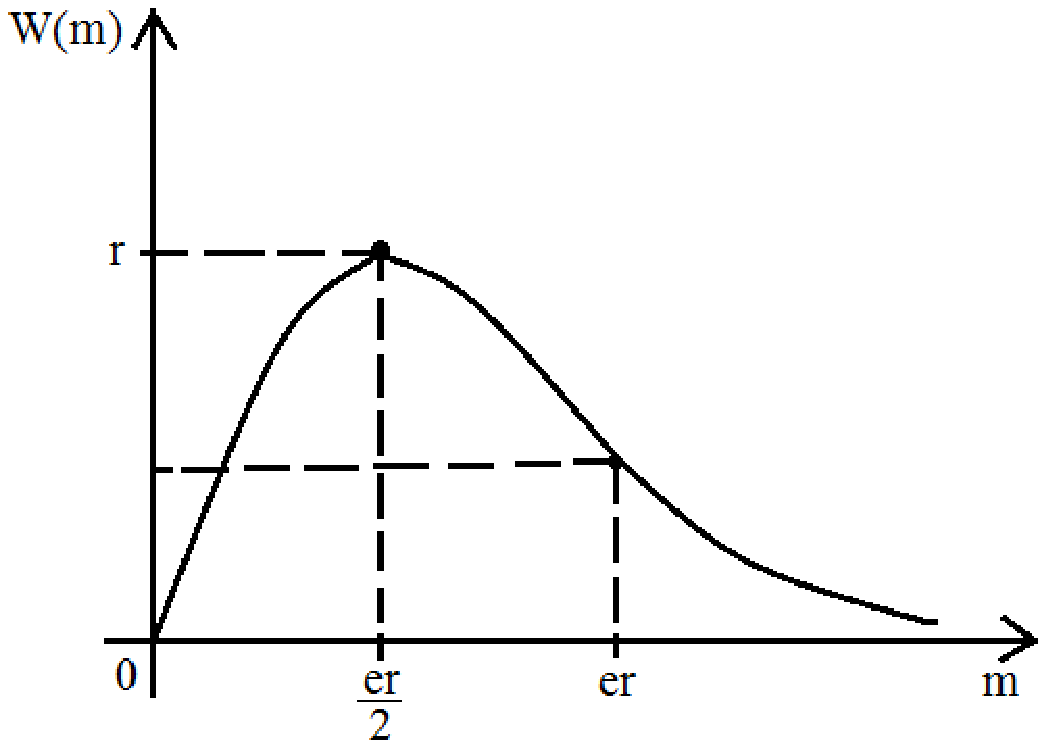}
\caption{Plot of the function $W(m)$ from Eq. (5.5), in terms of $m$ at constant $r$. It reaches its maximum value of $W = r$ at $m = er/2$, where the two side derivatives are different but finite.}
\label{68389Fig4}
\end{figure}

\end{document}